\documentclass[manuscript]{aastex}

\usepackage{graphicx}
\usepackage{amsfonts,amsmath,amssymb}
\usepackage{natbib}
\usepackage{hyperref}
\hypersetup{colorlinks=false,pdfborder={0 0 0},}
\usepackage{epstopdf}

\slugcomment{To appear in The Astrophysical Journal (ApJ)}

\shorttitle{The Bones of the Milky Way}
\shortauthors{Alyssa Goodman}

\begin{document}


\title{The Bones of the Milky Way}


\author
{Alyssa A. Goodman\affil{Harvard-Smithsonian Center for Astrophysics, Cambridge, MA 02138} 
Jo\~ao Alves\affil{University of Vienna, 1180 Vienna, Austria} 
Christopher N. Beaumont\affil{Harvard-Smithsonian Center for Astrophysics, Cambridge, MA 02138} 
Robert A. Benjamin\affil{University of Wisconsin-Whitewater, Whitewater, WI 53190}
Michelle A. Borkin\affil{Harvard University, Cambridge, MA 02138} 
Andreas Burkert\affil{University of Munich, Munich, Germany} 
Thomas M. Dame\affil{Smithsonian Astrophysical Observatory, Cambridge, MA 02138} 
James Jackson\affil{Boston University, Boston, MA 02215} 
Jens Kauffmann\affil{California Institute of Technology, Pasadena, CA 91125} 
Thomas Robitaille\affil{ Max Planck Institute for Astronomy, Heidelberg, Germany} 
Rowan J. Smith\affil{ Institut f\"ur Theoretische Astrophysik, Zentrum f\"ur Astronomie der Universi\"at Heidelberg, Heidelberg, Germany}
}

\begin{abstract}
The very long and thin infrared dark cloud ``Nessie" is even longer than had
been previously claimed, and an analysis of its Galactic location
suggests that it lies directly in the Milky Way's mid-plane, tracing out
a highly elongated bone-like feature within the prominent
Scutum-Centaurus spiral arm. Re-analysis of mid-infrared imagery from
the Spitzer Space Telescope shows that this IRDC is at least 2, and
possibly as many as 5 times longer than had originally been claimed by
Nessie's discoverers, \citet{Jackson2010}; its aspect ratio is therefore
at least 300:1, and possibly as large as 800:1. A careful accounting for
both the Sun's offset from the Galactic plane ($\sim 25$ pc) and the
Galactic center's offset from the $(l^{II},b^{II})=(0,0)$ position
shows that the latitude of the true Galactic
mid-plane at the 3.1 kpc distance to the Scutum-Centaurus Arm is not
$b=0$, but instead closer to $b=-0.4$, which is the latitude of Nessie
to within a few pc. An analysis of the radial velocities of low-density (CO) and
high-density (${\rm NH}_3$) gas associated with the Nessie dust feature
suggests that Nessie runs along the Scutum-Centaurus Arm in
position-position-velocity space, which means it likely forms a dense
`spine' of the arm in real space as well. The Scutum-Centaurus arm is
the closest major spiral arm to the Sun toward the inner Galaxy, and, at the longitude of Nessie, it is almost perpendicular to our line of sight, making Nessie the easiest feature to see as a shadow elongated along the Galactic Plane from our location.  Future high-resolution dust mapping and molecular line observations of the harder-to-find Galactic ``bones" should allow us to exploit the Sun's position above the plane to gain a (very foreshortened) view ``from above" of the Milky Way's structure.

\end{abstract}



\section{Introduction}

Determining the structure of the Milky Way, from our vantage point within it, is a perpetual challenge for astronomers. We know the Galaxy has spiral arms, but it remains unclear exactly how many \citep[cf.][]{Vallee2008a}. Recent observations of maser proper motions give unprecedented accuracy in determining the three-dimensional position of the Galaxy's center and rotation speed \citep{Reid2009,Brunthaler2011}. But, to date, we still do not have a definitive picture of the Milky Way's three dimensional structure.

The analysis offered in this paper suggests that some Infrared Dark Clouds\footnote {The term ``Infrared Dark Cloud" or ``IRDC" typically refers to any cloud which is opaque in the mid-infrared.}---in particular very long, very dark, clouds---appear to delineate major features of our Galaxy as would be seen from outside of it. In particular, we study a $>3^{\circ}$-long cloud associated with the IRDC called ``Nessie" \citep{Jackson2010}, and we show that it appears to lie parallel to, and no more than just a few pc from, the true Galactic Plane.

Our analysis uses diverse data sets, but it hinges on combining those data sets with a modern understanding of the meaning of Galactic coordinates. When, in 1959, the IAU established the current system of Galactic $(l,b)$ coordinates \citep{Blaauw1959}, the positions of the Sun with respect to the ``true" Galactic disk, and of the Galactic Center, were not as well determined as they are now. As a result, the Galactic Plane is typically \textbf{not} at $b=0$, as
projected onto the sky. The exact offset from $b=0$ depends on distance, as we explain in \S \ref{lookingdown}. Taking these offsets into account, one can profitably re-examine data relevant to the Milky Way's 3D structure.  The Sun's vantage point slightly ``above" the plane of the Milky Way offers useful perspective.

``IRDCs" are loosely defined as clouds with column densities high enough to be obvious as patches of significant extinction against the diffuse galactic background at mid--infrared wavelengths.  \citet{Peretto2009a} set the boundaries of IRDCs at an optical depth of 0.35 at $8~\rm{}\mu{}m$ wavelength, equivalent to an $\rm{}H_2$ column density $\approx{}10^{22}~\rm{}cm^{-2}$. In the \citet{peretto2010:irdcs-mass-density} sample, clouds have average column densities of a few $10^{22}~\rm{}cm^{-2}$. Some IRDCs actively form high--mass stars (e.g., \citealt{pillai2006:g11} and \citealt{rathborne2007:irdc-msf}). \citet{kauffmann2010:irdcs} explain that while some ``starless" IRDCs are potential sites of future high--mass star formation, and the few hundred densest and the most massive, IRDCs may very well contain a large fraction of the star--forming gas in the Milky Way, it is still true that most IRDCs are not massive and dense enough to form high--mass stars. Thus, a small number of very dense and massive IRDCs may be responsible for a large fraction of the galactic star formation rate, and an extragalactic observer of the Milky Way might ``see" IRDCs not unlike Nessie hosting young massive stars as the predominant mode of star formation here.  

The traditional ISM-based probes of the Milky Way's structure have been HI and CO. Emission in these tracers gives line intensity as a function of velocity, so the position-position-velocity data resulting from HI and CO observations can give three dimensional views of the Galaxy, if a rotation curve is used to translate line-of-sight velocity into a distance. Unfortunately, though, the Galaxy is filled with HI and CO, so it is very hard to disentangle features when they overlap in velocity along the line of sight. Nonetheless, much of the basic understanding of the Milky Way's spiral structure we have now comes from HI and CO observations of the Galaxy, much of it from the compilation of CO data presented by \citet{Dame2001}.

Recently, several groups have targeted high-mass star-forming regions in  the plane of the Milky Way for high-resolution observation. In their BeSSeL Survey, Reid et al. are using hundreds of hours of VLBA time to observe hundreds of regions for maser emission, which can give both distance and kinematic information for very high-density ($n>10^8$ cm$^{-3}$) gas \citep{Reid2009,Brunthaler2011}. In the HOPS Survey, hundreds of positions associated with the dense peaks of infrared dark clouds have now been surveyed for ${\rm NH}_3$ emission \citep{Purcell2012}, yielding high-spectral resolution velocity measurements towards gas whose density typically exceeds $10^4$ cm$^{-3}$. In follow-up spectral-line surveys to the ATLASGAL \citep{Beuther2012a} dust-based survey of the Galactic Plane, \citet{Wienen2012} have measured ${\rm NH}_3$ emission in nearly 1000 locations. The ThrUMMs Survey aims to map the entire fourth quadrant of the Milky Way in CO and higher-density tracers \citep{BarnesPeter2010}, and it should yield additional high-resolution velocity measurements.

Targets in high-resolution (e.g. BeSSeL) studies are usually identified based on continuum surveys, which show the locations of the highest column-density regions, either as extinction features (``dark clouds" in the optical, ``IRDCs" in the infrared), as dust emission features (in surveys of the thermal infrared), or as gas emission features (e.g. HII regions).

Great power lies in the careful combination of continuum and spectral-line data when one wants to understand the structure of the ISM in three-dimensions. Thus, there have already been several efforts to combine dust maps with spectral-line data, whose goal is often the assignment of more accurate distances to particular clouds or regions \citep[e.g.][]{Foster2012}.  These improved distances allow for more reliable conversion of  measured quantities (e.g. fluxes) to physical ones (e.g. mass).

In this study, our aim is to combine morphological information from large-scale mid-infrared continuum ``dust" maps of the Galactic Plane with spectral-line data, so as to understand the nature of very long infrared dark clouds that appear parallel to the Galactic Plane. We focus in particular on the IRDC named ``Nessie" in the study presented by \citet{Jackson2010}.  In that 2010 paper, Nessie is shown to be a highly elongated filamentary cloud (see Figure \ref{fig:FindingChart})  exhibiting the after-effects of a sausage instability that led to several massive-star-forming peaks spaced at regular intervals.
We extend the work of Jackson et al. by first by literally ``extending" the cloud, to a length of at least 3 degrees (\S \ref{longer}). In \S \ref{3D}, we show that a careful accounting for the modern measures of
the Sun's height off of the Galactic mid-plane and of the true position of the Galactic Center imply that Nessie lies not
just parallel to the Galactic Plane, but \emph{in} the Galactic Plane. We consider what velocity-resolved measures of the
material associated with Nessie tell us about its three-dimensional position in the Galaxy, and we conclude, in \S \ref{spine} that Nessie likely marks the ``spine" of the Scutum-Centaurus arm of the Milky Way in which it lies. In \S \ref{future}, we consider the likelihood of finding more ``Nessie-like" structures in the future, and of using them, in conjunction with the Sun's vantage point just above the mid-plane of the Milky Way, to map out the skeleton of our Galaxy.

\section{Nessie is Longer than We Thought}
\label{longer}
\begin{figure}[h!]
\begin{center}
\includegraphics[width=1\columnwidth]{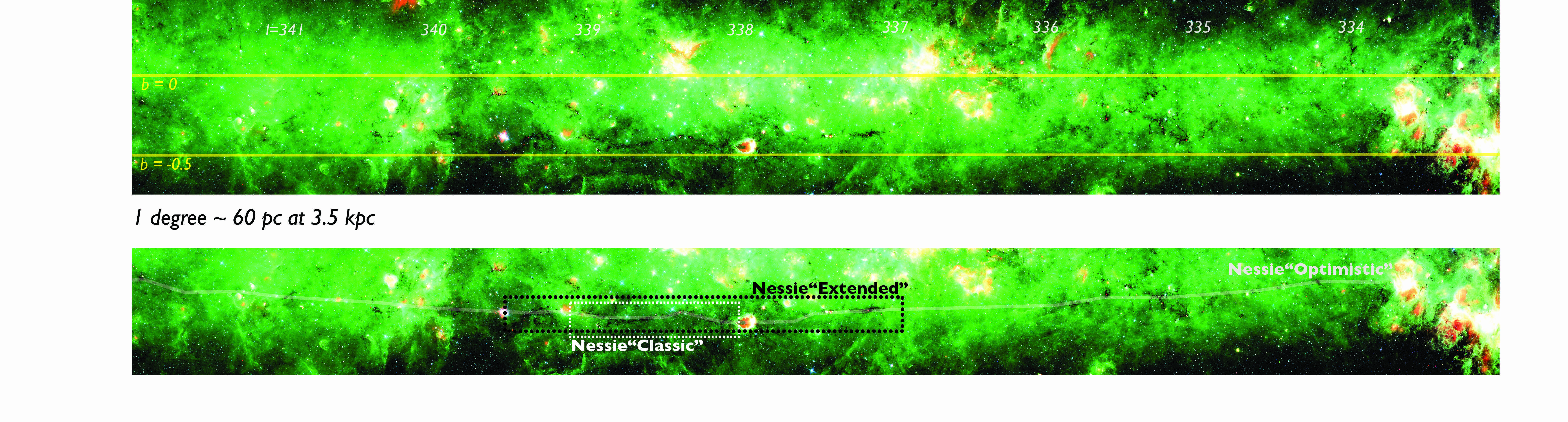}
\caption{\textbf{\label{fig:FindingChart}} Nessie ``Classic," ``Extended," and ``Optimistic."  It is recommended that this high resolution figure be viewed electronically, as zooming and panning are necessary to reveal the structures discussed.   Background imagery is a three-color image constructed from Spitzer where red shows MIPS 24 \micron, green IRAC 8.0 \micron, and blue IRAC 5.8 \micron. Image (brightness/contrast) adjustment used  emphasizes Nessie and also makes the image look a bit greener than is ``natural."  Yellow horizontal lines show $b=0^\circ$ and $b=-0.5^\circ$. Small white numbers show galactic longitude ($l$) in degrees. Full size version of this figure is available at \href{https://www.authorea.com/users/23/articles/249/master/file/figures/1nessie_findingchart/1nessie_findingchart_hires.jpg}{this link} (submit to Journal). A non-annotated high-dynamic-range view of this Spitzer image is available as a supplement to this paper at http://hdl.handle.net/1902.1/22050. (Note that related figures below show less exaggerated contrast, and zoom in on the central portions of Nessie.)}
\end{center}
\end{figure}

Nessie was discovered and named using Spitzer Space Telescope images that show the cloud as a very clear absorption feature at mid-infrared wavelengths \cite{Jackson2010}.  Using observations of the dense-gas tracer HNC, Jackson et al. (2010) further show that the section of the cloud from $l=$337.85 to 339.1 (labelled ``Nessie Classic" in Figure \ref{fig:FindingChart}) exhibits very similar line-of-sight velocities, ranging over $-40<v_{LSR}<-36$ km\ s$^{-1}$. The similarities of these line-of-sight velocities (cf.  Figure \ref{fig:HOPSoverlay}) is taken to mean that the cloud is a coherent, long structure, and not a chance
plane-of-the-sky projection of disconnected features. Thus, ``Nessie
Classic" is shown to be a dense,  long ($\sim 1^\circ$), narrow ($\sim 0.01^\circ$),
filament, and Jackson et al. (2010) ultimately conclude that it is undergoing a
sausage instability leading to density peaks hosting active sites of
massive star formation.

Our purpose in looking at Nessie again here is not to further analyze the star-forming nature of this cloud. Instead, our focus is on how long the full Nessie feature might be, and on what that length combined with Nessie's special position in the Galaxy  imply about how similar structures can be used to chart the structure of the Galaxy. Casual inspection of Spitzer imagery given in Figure \ref{fig:FindingChart} suggests that Nessie is at least two or three times longer than ``Nessie Classic," measuring at least $3^\circ$ long (``Nessie-Extended"). Very careful
inspection (pan and zoom Figure \ref{fig:FindingChart}) of the Spitzer images suggests that Nessie \emph{could be} even longer.  If
one optimistically connects what appear to be all the relevant pieces
then ``Nessie Optimistic" could be as much as $8^\circ$ long (light white chalk line in Figure \ref{fig:FindingChart}). The optimism involved in seeing the longest extent for Nessie could  be warranted if bright star-forming regions have broken up the continuous extinction feature, and/or if the background emission fluctuates enough to make the extinction  hard to detect.

\begin{figure}[h!]
\begin{center}
\plotone{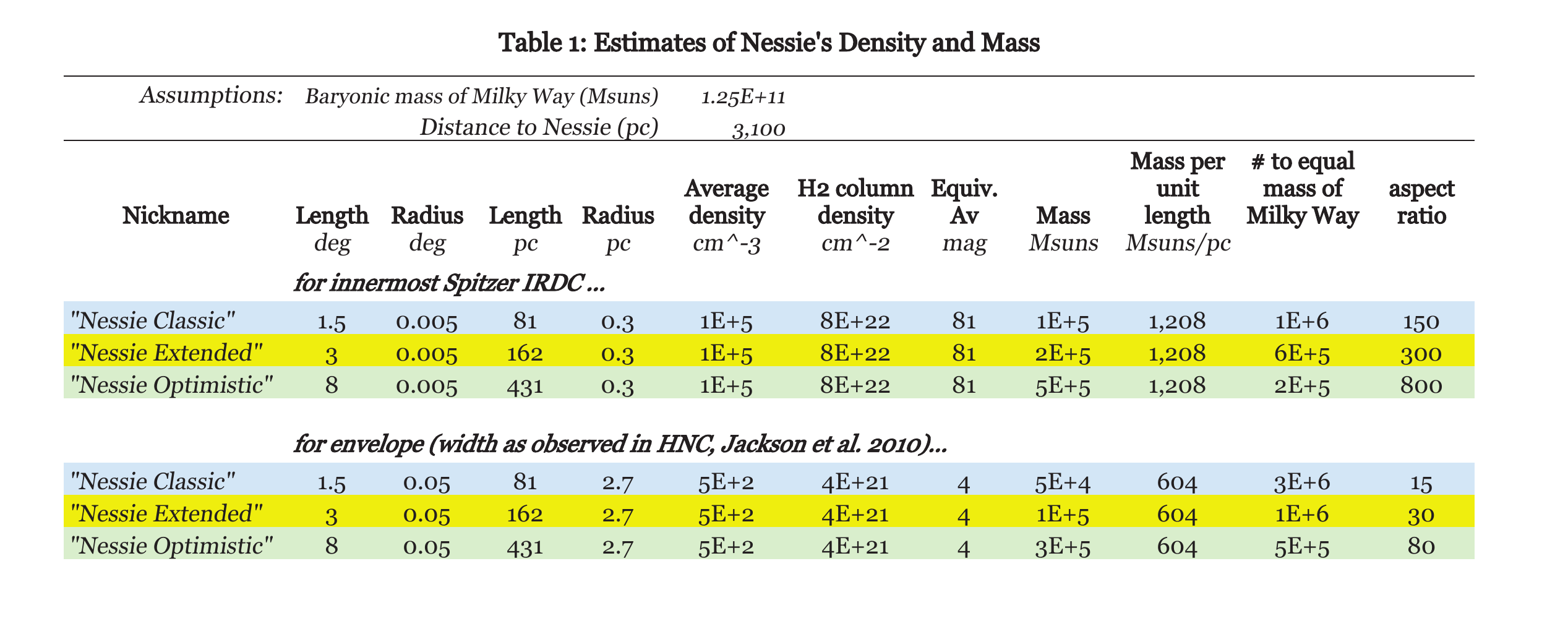}
\caption{\label{fig:table1} Estimates for the density and mass of Nessie, under various assumptions about its length.  The {\bf top set} of values shows estimates for a cylinder with radius, length, and average density appropriate to the Spitzer ``shadow" shown in Figure \ref{fig:FindingChart}.  The average density is set to $10^5$ cm$^{-3}$ so as to achieve a visual extinction of $\sim 100$ magnitudes, consistent with the typical extinction on tenths of pc scales in an IRDC.  The {\bf lower set} of values is different only in that a density of 500 cm$^{-3}$ is used, as an estimate of the typical density of the region seen to emit in HNC by Jackson et al. 2010. Readers can test alternative assumptions for the calculations shown here in the interactive version of this table at \href{http://tinyurl.com/nessietable}{http://tinyurl.com/nessietable}.}

\end{center}
\end{figure}

Determining the physical, three-dimensional, nature of extensions to the
Nessie cloud requires a detailed analysis of the velocity of the gas
associated with the dust responsible for mid-IR extinction. We
offer such an analysis below (\S\ref{3D}), but here we note that
if Nessie (as is nearly certain given its velocity range) lies in or
near the Scutum-Centaurus Arm of the Milky Way, then its distance is
roughly 3.1 kpc (cf. Jackson et al. 2010). At that distance, Nessie
Classic is roughly 80 pc long, Nessie Extended is 160 pc long, and
Nessie Optimistic is 430 pc long. For any of these lengths, the dark filament's
width is of order 0.01 degrees (0.5 pc), according to Jackson et al.'s (2010) analysis of the Spitzer imagery. Thus, clouds's axial ratio is about
150 for Nessie Classic, 300 for Nessie Extended, and nearly three times
more, 800, for Nessie Optimistic. (These calculations are based on Table 1 a publicly-available interactive spreadsheet, at
\href{https://docs.google.com/spreadsheet/ccc?key=0AhIRxiTe1u6BdDlXOC10Zkd3WUNNZHVnRlhfeWhJYlE}{this link}, a snapshot of which is shown as Figure \ref{fig:table1}.)

\section{The Three-Dimensional Position of Nessie within the Milky Way}
\label{3D}

\subsection{Looking ``Down" on the Galaxy}
\label{lookingdown}
Astronomers would love to leave the Milky Way, so that we could observe its spiral pattern face-on, as we do for other galaxies.  But our Sun is so entrenched in the Milky Way's plane that an ``overhead view" of the Milky Way's structure is impossible.  Or is it?  What if the Sun were just far enough above the Galactic Plane that we could use its height to give ourselves a tiny bit of perspective on the Galactic Plane that would spatially separate long skinny in-plane features located at different distances to be at different projected latitudes?  Turns out we {\it are} lucky in this way--the Sun {\it is} apparently located a bit above the Plane (see below), and we can use that vantage point to out advantage.  

To understand why most modern astronomers do not typically think about the possibility or value of an overhead view, we need to consider the origin of our current Galactic coordinate system, and our current understanding of the Sun's and the Galactic Center's 3D positions.  Writing in 1959 on behalf of the International Astronomical Union's (IAU's) sub-commission 33b, Blaauw et al.
wrote: 
\begin{quotation}
The equatorial plane of the new co-ordinate system must of
necessity pass through the sun. It is a fortunate circumstance that,
within the observational uncertainty, both the sun and Sagittarius A lie
in the mean plane of the Galaxy as determined from hydrogen
observations. If the sun had not been so placed, points in the mean
plane would not lie on the galactic equator.
\end{quotation}
In a further explanation of the IAU system in 1960, Blaauw et al. explain that stellar observations did, at that time, indicate  the Sun to be at $z_{\rm Sun}=22 \pm 2$ (22 pc above the plane), but the authors then discount those observations as too dangerously affected by hard-to-correct-for extinction in and near the Galactic Plane \citep{1960MNRAS.121..123B}.   Instead, the 1959 IAU system relies on the 1950s measurements of HI, which showed the Sun to be at $z_{\rm Sun}=4\pm 12$ pc off the Plane, consistent with the Sun being directly in the Plane ($z_{\rm Sun}=0$).  Interestingly, since the 1950s, the Milky Way's HI layer has been shown to have corrugations on the scale of 10's of pc \citep{Malhotra1995}, and there may be similar fluctuations in the mid-plane of the ${\rm H_2}$ \citep{Malhotra1994}, so it is still tricky to use gas measurements to determine the Sun's height off the plane. Although the Sun's ($\sim 25$
pc) offset from the Galactic plane is not large in comparison with the
half-thickness of the plane as traced by Population I objects such as
GMCs and HII regions ($\sim 200$ pc; \citet{2013A&ARv..21...61R}), it is much larger than the thickness of an extremely thin gas layer that may, as we argue in \S \ref{spine}, be traced out by Nessie-like ``bones" of the Milky Way. 

Astronomers today are still using the $(l^{II}, b^{II})$ Galactic coordinate system defined by
\citet{Blaauw1959}, but it is \emph{not} still the case, within observational uncertainty, that the
Sun is in the mean plane of the Galaxy, and the true position of the Galactic Center is no longer at $(l^{II}=0, b^{II}=0)$.
Instead, a variety of lines of evidence \citep{Chen2001b,Maiz-Apellaniz2001a,Juric2008a} show that the Sun is approximately 25 pc above the stellar Galactic mid-plane, and VLBA proper motion observations of masers show that the Galactic Center is about 7 pc below where the $(l^{II}, b^{II})$ system would put it, at $b=-0.046^\circ$ \citep{Reid2004}. After analysis of the spatial distribution of dense gas revealed in the ATLASGAL survey showed a mean at negative latitude, \citet{Beuther2012a} noted that their observations
\begin{quotation}
are indicative of a real global offset of the Galactic mid-plane from its conventional position where the axis between the Sun and the Galactic center is located at $b=0$ deg.
\end{quotation}
 The offsets of the Sun and the Galactic Center from their IAU ``$b=0$" positions, as anticipated by Blaauw et al., imply that ``points in the mean plane [do] not lie on the galactic equator.''

\begin{figure}[h!]
\begin{center}
\includegraphics[width=1\columnwidth]{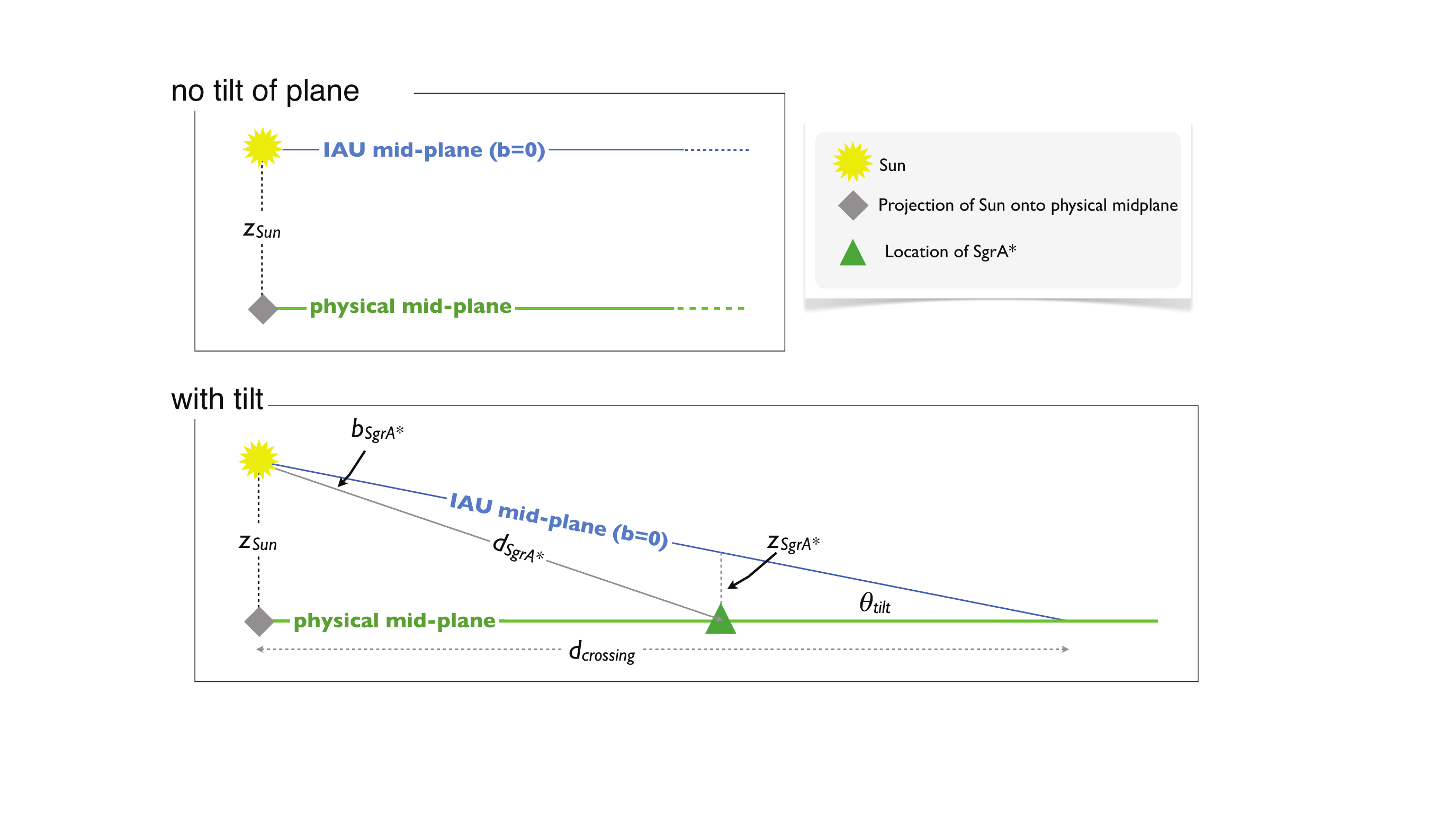}
\caption{\textbf{\label{fig:galcoords}}. Schematic side-views of the of the physical mid-plane of the Galaxy with respect to the IAU-defined mid-plane. The drawings are {\bf not to scale}. In both panels, $z_{\rm Sun}$ represents the height of the Sun above the Galactic mid-plane.  In the \textbf{upper panel}, and in figures labeled ``no tilt of plane" below, we consider \textit{only} the Sun's offset from the plane in calculating the observed coordinates of the true ``physical" mid-plane.  In the \textbf{lower panel}, and in figures labeled ``with tilt" below, we also take the offset, $z_{\rm SgrA*}$, of the Galactic Center (Sgr A*) into account.  For reference, if the distance to SgrA*, $d_{\rm SgrA*}$, is 8.5 kpc and  $z_{\rm Sun}$=25 pc and $z_{\rm SgrA*}=7$ pc (based on $b=-0.046^\circ$ for SgrA*) then the angle by which the IAU mid-plane is tilted with respect to the physical plane is $\theta_{\rm tilt}=0.12^\circ$ and the distance from the Sun to where the two planes cross is $d_{\rm crossing} =12$ kpc.}
\end{center}
\end{figure}

Figure \ref{fig:galcoords} shows a schematic (not-to-scale) diagram of the effect of the Sun's and the Galactic Center's offsets from the  mid-plane defined by the IAU in 1959 (and still in use as $(l^{II}, b^{II})$ today).  The tilt of the the true, physical, Galactic mid-plane to the presently IAU-defined plane means that, within about 12 kpc of Sun\footnote{12 kpc is the approximate distance where the physical and IAU planes cross, on a line toward the Galactic Center.  Along other directions toward the Inner Galaxy, as shown in the lower panel of Figure \ref{fig:topview}, it will be further to the crossing point, and toward the Outer Galaxy, for a ``flat" disk, the mid-plane will always appear at negative latitudes.} any feature that is truly ``in" the Galactic mid-plane will appear on the Sky at negative $b^{II}$.  Figure \ref{fig:coloredlines} shows an example of this effect, where the rainbow-colored dashed line indicates the sky position of the physical Galactic mid-plane at a Nessie-like distance of 3.1 kpc (assuming the the Sun is 25 pc off the plane, a distance to Sgr A** of 8.5 kpc, a rotation speed for the Milky Way of 220 km\ s$^{-1}$, and (U,V,W) motion for the Sun of 11.1, 12.4, and 7.2 km\ s$^{-1}$, respectively).

\subsection{Using Rotation Curves and Velocity Measurements to Place Nessie in 3D}

\begin{figure}[!htbp]
\begin{center}
\includegraphics[width=0.6\columnwidth]{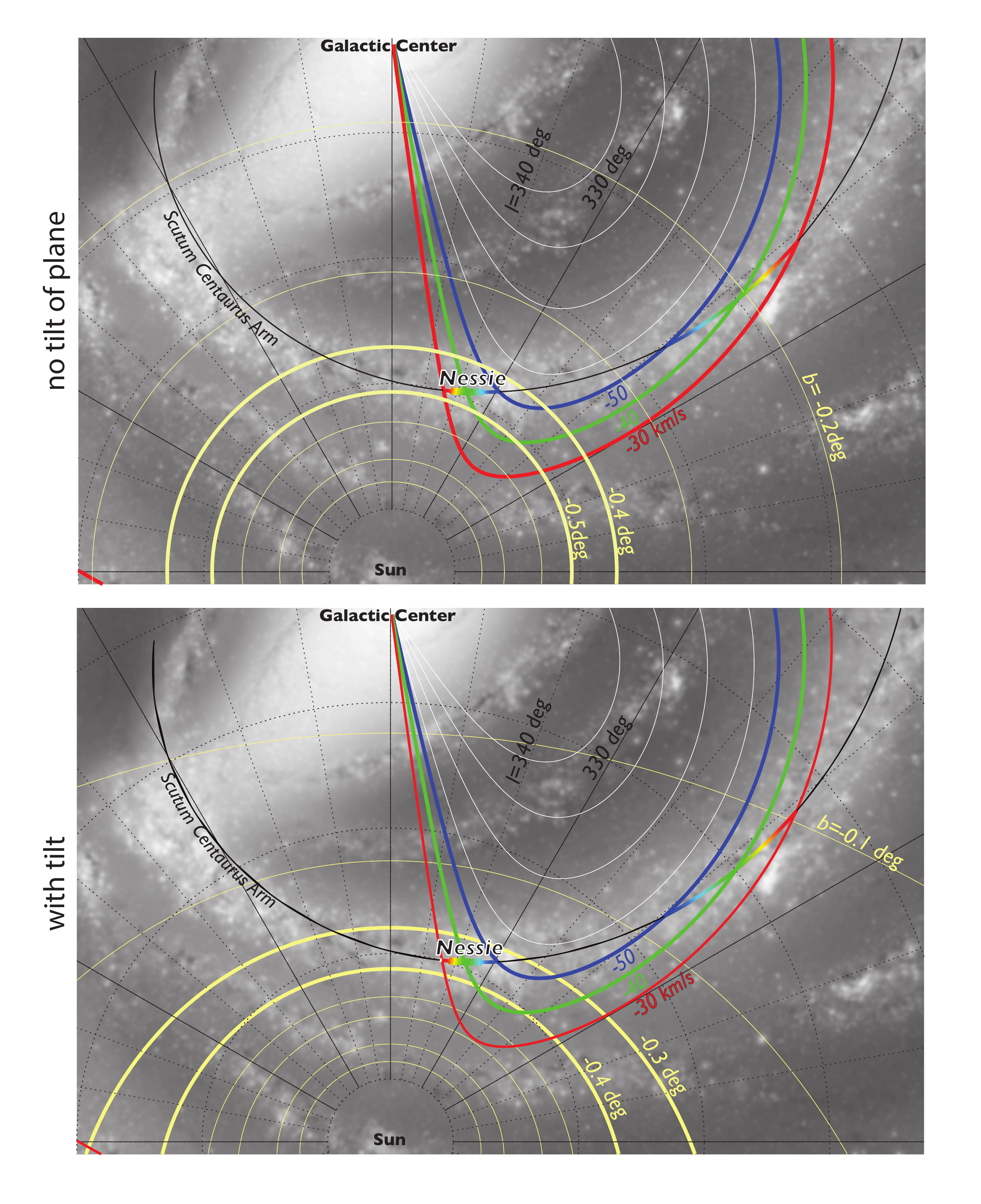}
\caption{\textbf{\label{fig:topview}} For the fourth quadrant of the Milky Way, contours of constant LSR velocity of -30,-40,and-50 km s$^{-1}$ \citep[using a rotation curve from][ ]{McClureGriffiths2007} are superimposed on a cartoon model of the Milky Way. CO and dense gas observations (see below) give LSR velocities associated with Nessie (shown here as rainbow curve) near -40 km$^{-1}$, placing Nessie in the Scutum-Centaurus arm (highlighted in black), about 3.1 kpc from the Sun. Yellow-highlighted curves show positions in the Galaxy that would have the labeled value of Galactic Latitude ($b$) when viewed from the Sun.  In the \textbf{top panel}, \textit{only} the height of the Sun off the plane (taken to be 25 pc in this example) is considered in drawing the iso-b curves, and in the \textbf{bottom panel}, a 7 pc offset of the Galactic Center (see text), which causes an overall tilt of $0.12^\circ$ is \textit{also} taken into account. }
\end{center}
\end{figure}

Ever since velocity-resolved observations of stars and gas have been possible, astronomers have been modeling the rotation pattern of the Milky Way.   Using a measured rotation curve for the Milky Way's gas \citep[e.g.][]{McClureGriffiths2007}, one can translate observed LSR velocities to a unique distance in the Outer Galaxy, and to one of two possible (``Near" or ``Far") distances toward the Inner Galaxy.   Figure \ref{fig:topview} shows iso-$v_{LSR}$ contours toward the Inner Galaxy, around the longitude range of Nessie, superimposed on the data-driven cartoon of our current understanding of the Milky Way's structure.  Notice that velocities associated with the near-side of the Scutum-Centaurus Arm in Nessie's longitude range should be near 40 km\ s$^{-1}$.

\begin{figure}[!htb]
\begin{center}
\includegraphics[width=1\columnwidth]{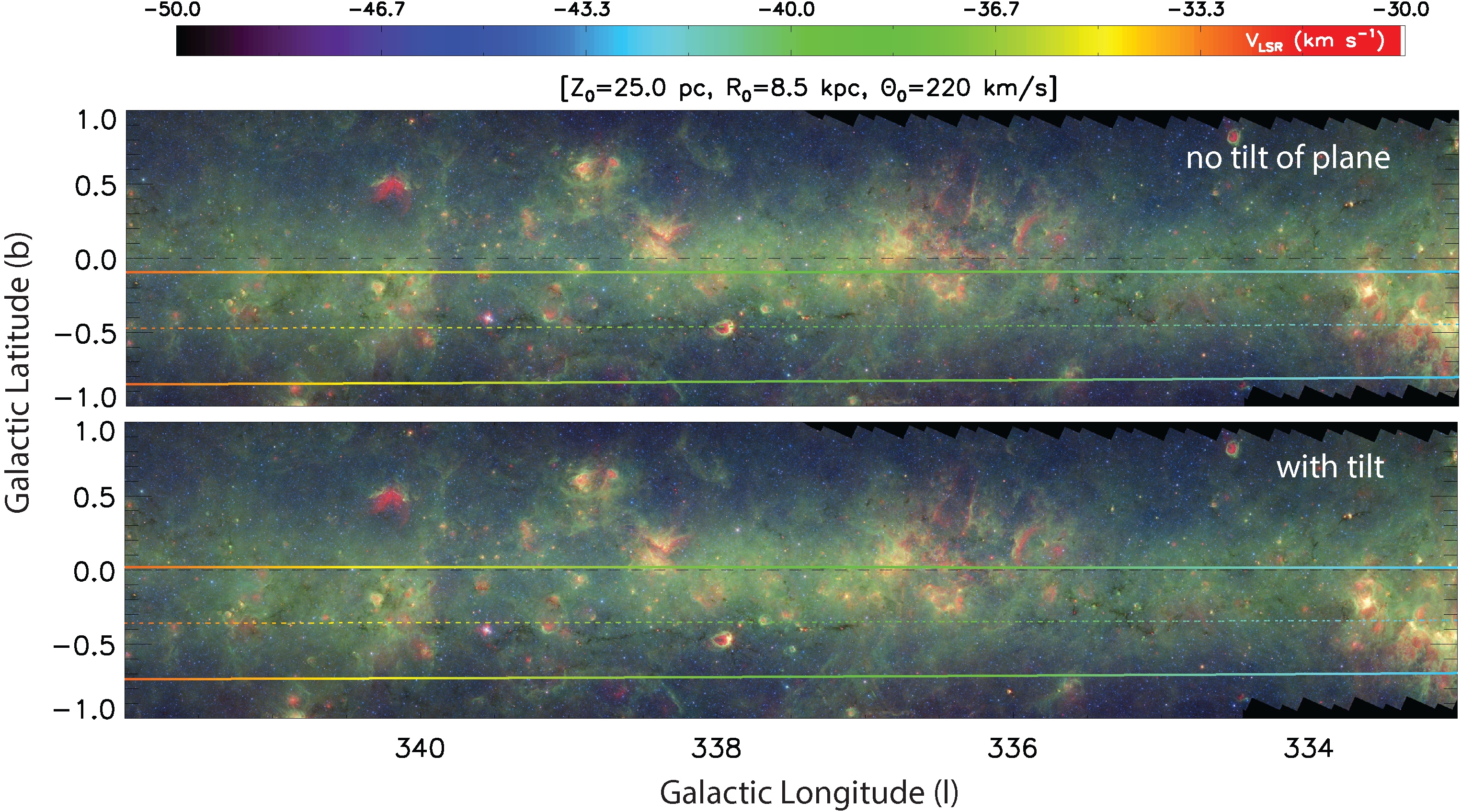}
\caption{\textbf{\label{fig:coloredlines}}. Predicted Sky Position and Radial Velocities for the Scutum-Centaurus Arm of the Milky Way.  In both panels, the colored lines are color-coded by velocity, given in the color bar at the top.  The colored dashed line shows the predicted position for the Galactic Plane for the near side of the Scutum-Centaurus Arm (roughly 3.1 kpc from the Sun).  The solid colored lines show $\pm 20$ pc from the mid-Plane at the same ($\sim3.1$ kpc) distance.   These lines and colors are calculated using a model of a (flat) Milky Way described by the parameters shown below the color bar.  Background imagery is from Spitzer, as in Figure \ref{fig:FindingChart}. A black dashed line emphasizes the position of $b=0$.  As in Figure \ref{fig:topview}: in the \textbf{top panel}, only a 25 pc offset of the Sun above is taken into account; and in the \textbf{bottom panel}, a 7 pc offset for the Galactic Center is also used in the calculation.}
\end{center}
\end{figure}

Combining a modern estimate for the Sun's height above the plane ($z_{\rm Sun}\sim 25$ pc), with the IAU Galactic coordinate definitions, we can determine where the physical mid-Plane of the Galaxy \textit{should} appear in the $(l^{II}, b^{II})$ system at any particular distance from the Sun.  Figure \ref{fig:coloredlines} shows where the Scutum-Centaurus Arm would appear on the Sky (for a distance to Sgr A** of 8.5 kpc, a rotation speed for the Milky Way of 220 km\ s$^{-1}$, and (U,V,W) motion for the Sun of 11.1, 12.4, and 7.2 km\ s$^{-1}$, respectively).   As its caption explains in detail, Figure \ref{fig:coloredlines}'s colored lines are associated with the near part of the Scutum-Centaurus Arm. Two versions of this plane-of-the-Sky view are shown, one \textit{only} accounting for the offset of the Sun, and the other also accounting for the tilt of the coordinate system caused by the Galactic Center also not lying in the IAU plane (see Figure \ref{fig:galcoords}). 

The dashed colored lines in Figure \ref{fig:coloredlines}, indicating the predicted position of the Galactic Plane on the Sky at the distance to the near side of the Scutum-Centaurus Arm, pass almost directly through Nessie, regardless of whether or not one considers the ``tilt" of the coordinate system caused by SgrA*'s offset.  Solid colored lines show 20 pc above and below the Plane at the distance to the Scutum-Centaurus Arm, so Figure \ref{fig:coloredlines} makes it is very clear that Nessie lies within just a few pc of the Plane, along its entire length. This is either an extremely fortuitous coincidence, or an indication that Nessie is tracing a significant feature that effectively marks the mean location of the Galactic Plane.  Given the waviness of the plane on 10 pc scales (see above, \citep{Malhotra1994}), the location at even less than 10 pc from the mean plane may be fortuitous--but the location so close to the mean is not. 

\subsubsection{CO Velocities}
\label{CO}
\begin{figure}[!htb]
\begin{center}
\includegraphics[width=1\columnwidth]{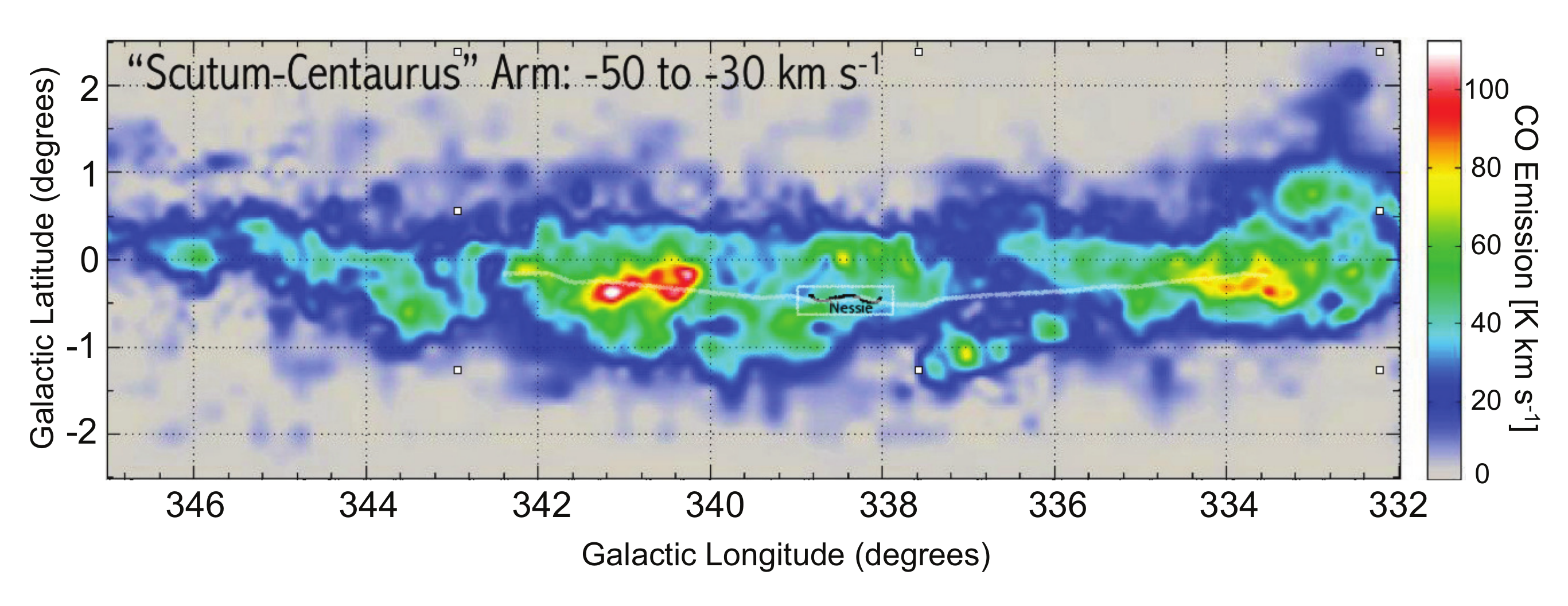}
\caption{{\label{fig:COarm}} CO emission, integrated between -50 and -30 km s$^{-1}$, as projected on the sky, based on data from \citep{Dame2001}, with trace of Nessie Optimistic (light white line) from Figure \ref{fig:FindingChart} superimposed. The dark black squiggle labeled ``Nessie" (surrounded by a light white box) marks the position of Nessie Classic.}
\end{center}
\end{figure}

CO observations trace gas with mean density around 100 cm$^{-3}$.   CO emission associated with the Scutum-Centaurus Arm of the Milky Way \citep{Dame2001,Dame2011} is shown in Figure \ref{fig:COarm}, which presents a plane-of-the-sky map integrated over  $-50 <v_{LSR}< -30$ km\ s$^{-1}$.  The velocity range is centered on -40 km\ s$^{-1}$, the average velocity of the Scutum-Centaurus Arm in Nessie's longitude range (see Figures \ref{fig:topview} and \ref{fig:coloredlines}).  The white chalk line superimposed on Figure \ref{fig:COarm} is the same tracing of ``Nessie Optimistic" shown in Figure \ref{fig:FindingChart}.  The black feature labeled ``Nessie" refers to ``Nessie Classic."   

Judging by-eye, the vertical (latitude) centroid of the CO emission in Figure \ref{fig:COarm} appears to follow Nessie remarkably well, even out to the full $8^\circ$ (430 pc) extent of Nessie Optimistic.  We have also calculated a curve representing the locus of latitude centroids for CO in this velocity range, and even at this coarse resolution, a curve following Nessie's shape is clearly a better fit than a straight line passing through the CO centroids. 

Table 1 estimates that the Nessie IRDC has a typical ${\rm H_2}$ column density of $\sim 10^{23}$ cm$^{-2}$ and a typical volume density of $\sim 10^5$ cm$^{-3}$.    Thus, the plane-of-the-sky coincidence of the line-of-sight-velocity-selected ``Scutum-Centaurus" CO emission  and the mid-IR extinction suggests that the Nessie IRDC may be a kind of dense ``spine" or ``bone" of this section of the Scutum-Centaurus Arm, as traced by much-less-dense ($\sim 100$ cm$^{-3}$) CO-traced gas.  But, the spatial resolution of the CO map is too low ($8'$), and the 20 km\ s$^{-1}$ velocity range associated with the Arm in CO is too broad to decide based on this evidence alone whether Nessie is a well-centered ``spine" or just a long skinny feature associated with, but potentially significantly inclined to, the Scutum-Centaurus Arm. 

\subsubsection{NH$_3$ Velocities}
\label{ammonia}
To estimate the 3D orientation of Nessie more precisely, we need to employ a gas tracer whose emission is sparser than CO's, and more associated with high-density gas, in position-position-velocity space. Many recent studies have shown that IRDCs typically host over-dense blobs of gas (often called ``clumps" or ``cores") that provide the gaseous reservoirs for the formation of massive stars.  Thus, several studies have been undertaken to survey IRDCs and their ilk for emission in molecular lines that trace high-density ($\gg 10^3$ cm$^{-3}$), potentially star-forming, gas.  

The H$_2$O Southern Galactic Plane (HOPS) Survey \citep{Purcell2012} has surveyed hundreds of sites of massive star formation visible from the Southern Hemisphere for ${\rm NH}_3$ emission, which traces gas at densities $n\gtrsim 10^4$ cm$^{-3}$.   The HOPS targets were selected based on H$_2$O maser emission, thermal molecular emission, and radio recombination lines, so as to include nearly all known regions of massive star formation within the surveyed area.  These ``massive-star-forming region" selection criteria mean that the HOPS database includes ${\rm NH}_3$ spectra for dozens of positions within the longitude range covered by Nessie.   

\begin{figure}[!htb]
\begin{center}
\includegraphics[width=1\columnwidth]{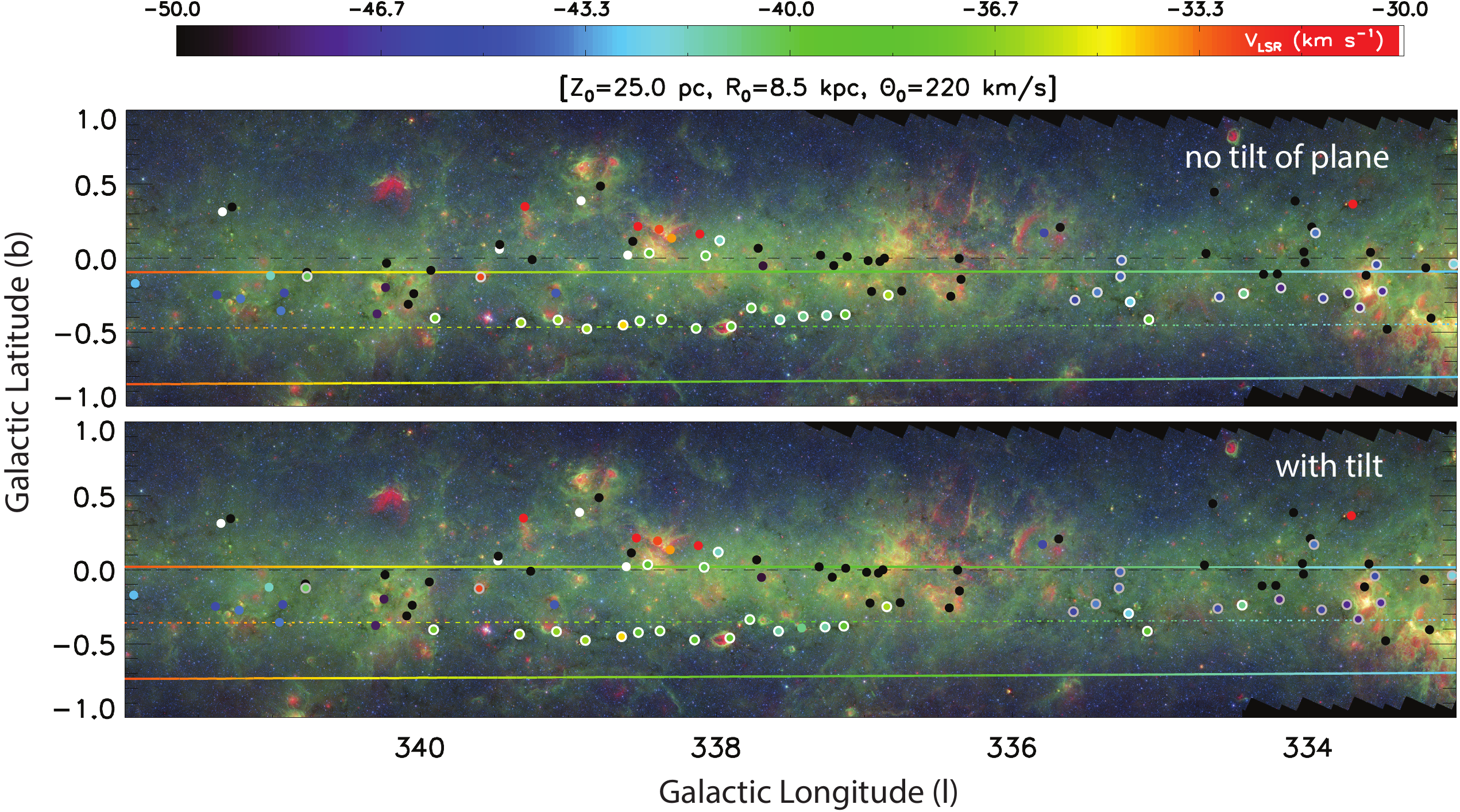}
\caption{{\label{fig:HOPSoverlay}} Superposition of HOPS Sources, color-coded by ${\rm NH}_3$-determined LSR velocity.  Colored lines have the same meaning (predicted LSR velocity) as in Figure \ref{fig:coloredlines}, so the colorful dashed lines at $b\sim-0.5^\circ$, in both panels, represent the physical Galactic mid-plane, and the solid colorful lines indicate 20 pc above and below the plane. Agreement of the NH$_3$ and predicted LSR velocity (color) to within 2.5 km s$^{{-}1}$ is indicated by a \textit{white circle} around the HOPS symbol, and \textit{grey} circles indicate agreement to within 5 km s$^{{-}1}$. As in Figures \ref{fig:topview} and \ref{fig:coloredlines}: in the \textbf{top panel}, only a 25 pc offset of the Sun above is taken into account; and in the \textbf{bottom panel}, a 7 pc offset for the Galactic Center is also used in the calculations.}
\end{center}
\end{figure}

\begin{figure}[!htbp]
\begin{center}
\includegraphics[width=0.6\columnwidth]{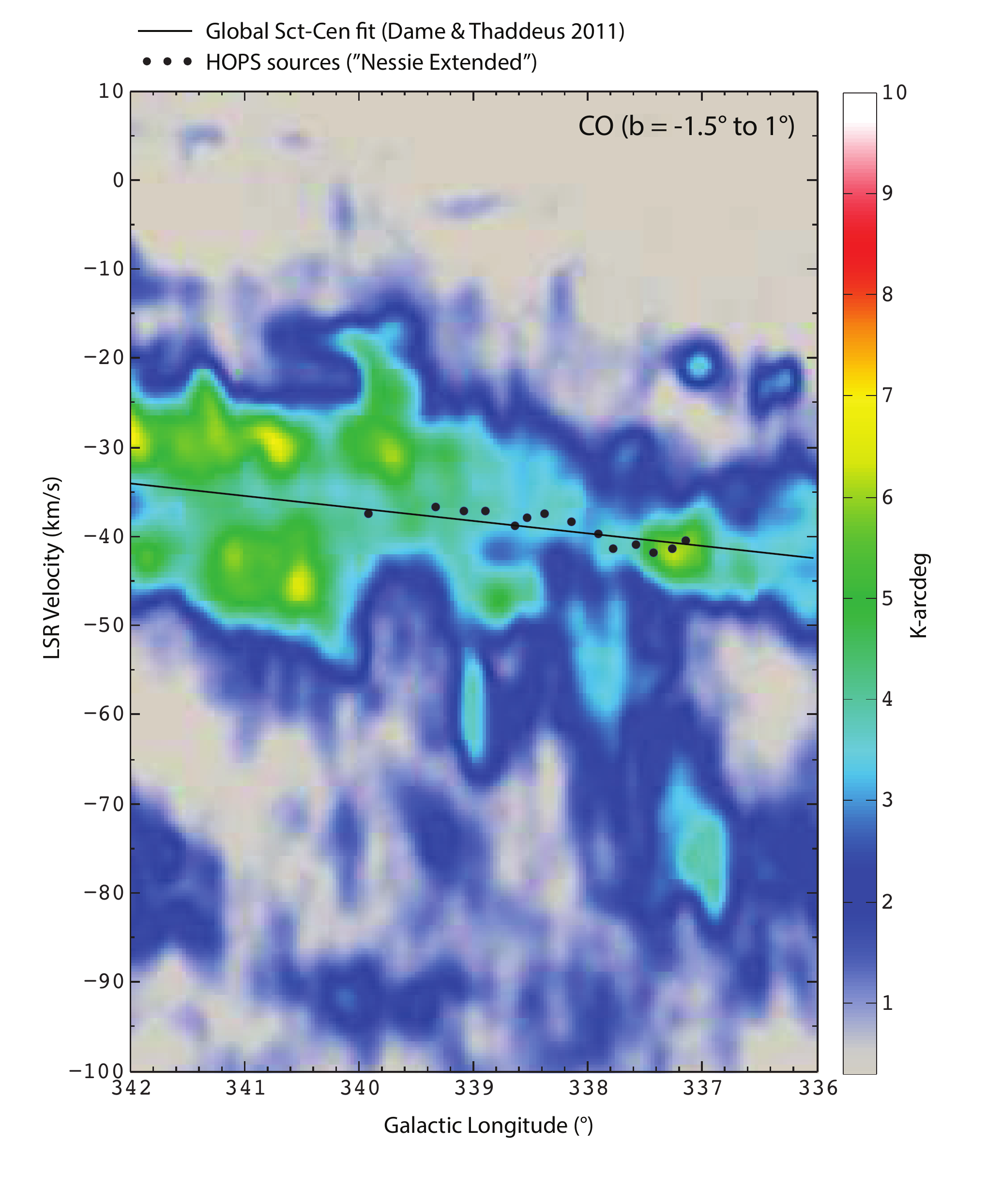}
\caption{{\label{fig:pvdiagram}} Position-velocity diagram of CO and ${\rm NH}_3$.  Colored background shows CO emission integrated over $-1.5<b<1^\circ$.  Black dots show HOPS sources coincident with Nessie Extended, also shown in Figure \ref{fig:HOPSoverlay} as white-circle-highlighted colored points.  The dots are plotted at the longitudes given in Figure \ref{fig:HOPSoverlay}, and the LSR velocities given by the centroid of the ${\rm NH}_3$ emission for each HOPS source. The black line shown is {\it not} a fit to the HOPS or the CO data shown: it is a segment of a {\it global} log-spiral fit to CO data for the entire Scutum-Centaurus Arm, extending almost $360^{\circ}$ around the Galaxy (taken from Fig. 4 of \cite{Dame2011}. }
\end{center}
\end{figure}

Figure \ref{fig:HOPSoverlay} adds an overlay of HOPS sources' ${\rm NH}_3$-determined LSR velocities to the information presented in Figure \ref{fig:coloredlines}, which illustrates how well Nessie fits the prediction of where the Scutum-Centaurus' arm's center would be in position-position-velocity space.  The (color-coded) velocities of the HOPS sources, for both Nessie Classic, and Nessie Extended (see Figure \ref{fig:FindingChart}), agree remarkably well with what is predicted for the Scutum-Centaurus Arm  (color-coded lines). Note that agreement of the  ${\rm NH}_3$ and predicted velocity to better than 5 km s$^{-1}$ is indicated by grey circles around the HOPS symbol.  White circles, indicating agreement to within 2.5 km s$^{{-}1}$,  surround all of the Nessie Extended sources, which are also shown in Figure \ref{fig:pvdiagram}, below.   Most of the HOPS sources coincident with Nessie Optimistic, especially at the lower longitudes, also appear likely to be associated with the Scutum-Centaurus arm, based on their agreement with arm velocities to within 5 km s$^{-1}$, as indicated by grey circles.   The velocities of sources at latitudes much different from Nessie's within this longitude range largely do {\it not} agree, and those sources are unlikely to be associated with the near-side of the Scutum-Centaurus Arm.     For reference, the typical velocity dispersion within molecular gas (with density $\sim 50$ cm$^{-3}$) is about 5 km s$^{-1}$ \citep{Larson1981}, so any white or grey outlined point in Figure \ref{fig:HOPSoverlay} is plausibly  associated with a structure whose systemic velocity is typical of the near side of the Scutum-Centaurus Arm.

For Nessie Classic, Jackson et al. (2010) had already noted a very narrow velocity range for dense gas associated with the IRDC, based on HNC observations.   What is new here is the three-dimensional (latitude, longitude, \textit{and} velocity) association of a {\it longer} Nessie's dense gas with predictions for where the centroid of the Milky Way's Scutum-Centaurus Arm's ``middle" would lie.  

Figure \ref{fig:pvdiagram}, which offers a position-velocity diagram of CO (color) and ${\rm NH}_3$ emission (black dots) together, shows the association of the Nessie-HOPS sources with the Scutum Centaurus Arm most clearly.  What is most remarkable about Figure \ref{fig:pvdiagram} is that the black line sloping through the figure is {\it not} a fit to the black dots representing the HOPS sources.  Instead, that line indicates the position-velocity trace of the Scutum-Centaurus Arm based on \citep{Dame2011} data for the full Galaxy, not just this small longitude range.   Figure \ref{fig:pvdiagram} implies that Nessie goes right down the ``spine" of the Scutum-Centaurus Arm, as best we can measure its position in CO position-velocity space.

\section{What is the Significance of Nessie-like structures within a Spiral Galaxy?}
\subsection{A Bone of the Galaxy}
\label{spine}
All the evidence presented in this paper, taken together, strongly suggests that Nessie forms a bone-like feature that closely follows the center of the Scutum-Centaurus Arm of the Milky Way.  How did it get there?  Is it the crest of a classic spiral density wave \citep{Lin1964}, or does it have some other cause?  One would na\"ievely expect that any feature this long and skinny that is not controlled by Galactic-scale forces will be subject to a variety of instabilities, and would not last long. 

Until very recently, no numerical simulation of gas in a Milky Way-like galaxy had sufficient resolution to reveal features as thin as Nessie.  As of 2013, state-of-the-art simulations by  \citet{2013MNRAS.432..653D} showed many elongated features along and between arms,\footnote{Simulation video online at  \href{http://empslocal.ex.ac.uk/people/staff/cld214/movies.html}{http://empslocal.ex.ac.uk/people/staff/cld214/movies.htm.}} but the thickness of the features was $\sim 10$ pc, an order of magnitude too coarse.

\begin{figure}[!htb]
\begin{center}
\includegraphics[width=0.9\columnwidth]{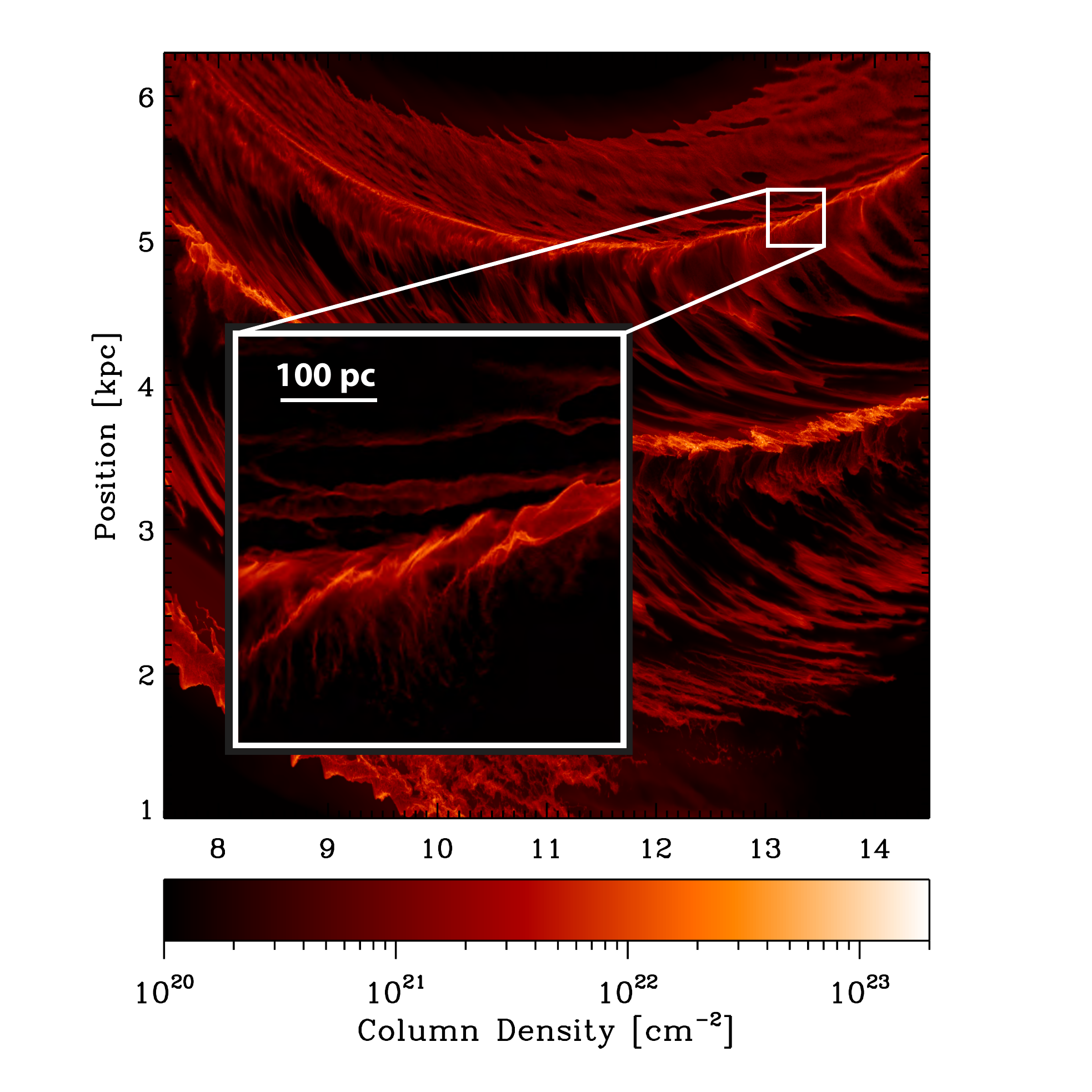}
\caption{{\label{fig:simulationtop}} Top-down view of the total column density in a simulation of structures forming within a spiral galaxy based on \citet{Smith2014}.  The main bright curving features show spiral arm structures.   Note the 100 pc scale bar within the zoom box, which shows that the simulation's resolution is very close to high enough to compare directly with structures with ``Nessie," whose width is just under 1 pc. }
\end{center}
\end{figure}

\begin{figure}[!htb]
\begin{center}
\includegraphics[width=0.9\columnwidth]{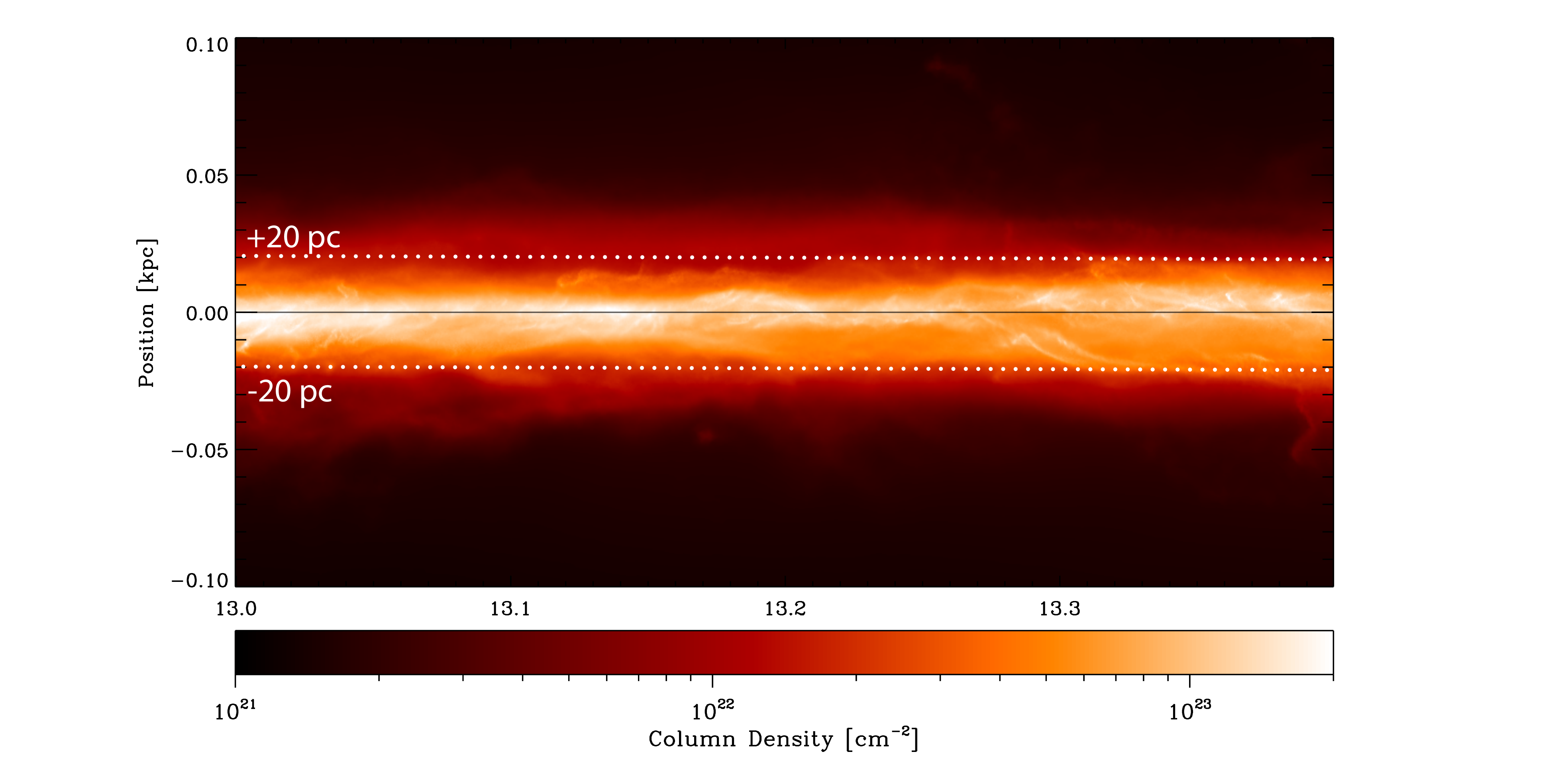}
\caption{{\label{fig:simulationedge}} Edge-on view of the total column density in a simulation of structures forming within a spiral galaxy based on \citet{Smith2014}.     Note the white dotted lines  drawn at 20 pc above and below the Galactic Plane (black line), and, by comparing with  Figure \ref{fig:coloredlines}, notice how the scale of the highly-elongated whitest high-density features in this simulation is approximately comparable to Nessie.}
\end{center}
\end{figure}

Figures \ref{fig:simulationtop} and \ref{fig:simulationedge} offer top-down and edge-on snapshots of a brand-new numerical simulation of the gas in a Milky Way-like spiral galaxy from \citet{Smith2014}, and this simulation is the {\it first} to offer resolution high enough to see potentially Nessie-like features. The AREPO moving mesh code used in these new simulations provides resolution $\sim 0.3$ pc resolution in regions where the gas density $n>10^3$ cm$^{-3}$, and slightly better in the highest-density cells. The high-resolution views of the simulation in the inset in Figure \ref{fig:simulationtop} and in Figure \ref{fig:simulationedge} exhibit many spatial features that are as long and thin as Nessie.  The overlain lines in Figure \ref{fig:simulationedge}  additionally show that the Nessie-like high-contrast filamentary structures like centered in the "Galactic Plane" of the simulation, and that they have vertical extent not unlike Nessie (cf. Figure \ref{fig:coloredlines}).  So as not to over-emphasize the apparently remarkable agreement of the \citet{Smith2014} simulations and Nessie-like structure, we should point out that the simulations do not include the results of feedback from stars and HII Regions, nor do they include magnetic fields.  Including these effects in future simulations will likely cause disruption of some filaments, and changes in topology and/or filament orientation where magnetic energy is significant.   For reference, the half-thickness of the Milky Way's plane as traced by Population I objects such as GMCs and HII regions is estimated to be $\sim 200$ pc \citep{2013A&ARv..21...61R}, which is the starting thickness of the evolving torus in the \citet{Smith2014} simulation. For now, we offer these simulations as the first evidence that gas in a spiral galaxy is very likely to contain very high-contrast, extraordinarily long and thin, filamentary structure marking out what we call the galaxy's ``skeleton."

Many of the features that appear to be trailing off from the major spiral arms apparent in Figure \ref{fig:simulationtop} are similar to the ``spurs" and ``feathers" that have been previously simulated and observed by E. Ostriker and colleagues \citep{Shetty2006,Vigne2008,Corder2008} and seen in the simulations of  \citet{2013MNRAS.432..653D}.   Figure \ref{fig:IC342} shows a recent infrared (WISE) image of the galaxy IC342 \citep{Jarrett2013}, and it is clear from that image that some `spiral' galaxies also exhibit inter-arm filaments that are even more pronounced than the simulated spurs and feathers.  Thus, both simulations and observations show long, filamentary, structures within, and between arms.  

The highest-resolution simulations \citep{Smith2014}  suggest that the highest-contrast filaments tend to largely be found {\it in} the plane of a spiral galaxy (Figure \ref{fig:simulationedge}) associated with arms, rather than inter-arm regions (Figure \ref{fig:simulationtop}).  But, high-density contrast alone does not imply that a filament lies along a spiral arm rather than between arms. The existence of many dense cores within Nessie (see \S \ref{ammonia}) implies an over-density of at least $10^3$ (comparing NH$_3$ gas and  CO-emitting gas) at many positions along the filament. Similar over-densities, and strings of dense cores, are seen in many IRDCs. \citet{Battersby} identified G32.02+0.06 as a sinuous, long, thin, 80-pc long ``Massive Molecular Filament" near (but not quite in) the Galactic plane, and in very recent work, \citet{Ragan2014}  added seven more ``Giant Molecular Filaments" almost as long and thin as Nessie to the list of very long-thin $\sim 10^5$ M$_\odot$ filamentary IRDCs known in the Milky Way.

In using observations to judge whether a specific highly-elongated filamentary cloud lies along (within) an arm, or is highly inclined to an arm, the velocity of the associated gas offers the most relevant evidence. In the case of Nessie, the analysis of the HOPS NH$_3$ data presented in \S \ref{ammonia}, showing agreement with spiral arm line-of-sight velocities, and its location at what is predicted to be the vertical center of the Scutum-Centaurus arm, argues very strongly that Nessie is, three-dimensionally, a spine-like ``bone" of the Scutum-Centaurus Arm. 

Estimates for the mass of Nessie under various assumptions are given in Table 1.  Jackson et  al. 2010  model Nessie as a(n unmagnetized) self-gravitating fluid cylinder supported against collapse by a ``turbulent" analog of thermal pressure, undergoing the sausage instability discussed in \citet{1953ApJ...118..116C}. For the observed line width of HNC, the theoretical critical mass per unit length, $m_l$, is 525 ${\rm\ M}_\odot\  {\rm pc}^{-1}$ for Nessie.  But, if, as Jackson et al. explain, one estimates $m_l$ using HNC emission itself and (uncertain) abundance values for HNC, then $110 < m_l < 5 \times 10^4 {\rm\ M}_\odot\  {\rm pc}^{-1}$. Given that the low end of this range ($110 {\rm\ M}_\odot\  {\rm pc}^{-1}$, favored by Jackson et al.) gives a very low value for extinction toward Nessie ($A_V \sim 4$ mag), we favor higher values of $m_l$, needed to be consistent with the observed IR extinction. Recent LABOCA observations of dust continuum emission from pieces of Nessie (Kauffmann, private communication), suggest that $m_l$ $\gtrsim$ $10^3$ in the mid-IR-opaque portions of Nessie.  
So, at present, it would appear that there is at least an order-of-magnitude uncertainty in $m_l$.  Some of this uncertainty is caused by the definition of Nessie's shape, which makes it unclear which ``mass" to measure in calculating $m_l$, but more is due to the vagaries of converting molecular line emission and/or dust continuum to true masses. 

What most interests us here is not Nessie's exact mass, mass per unit length, or density profile, but instead its total mass as a fraction of the Galaxy's mass, so that we can use that estimate in assessing how many Nessie-like structures may be findable in the Milky Way. To begin that assessment, we only need estimate Nessie's {\it total} mass.  Toward that end, Table 1 offers very simple estimates of the mass of cylinders, whose (constant) average density is set so that the typical extinctions associated with Nessie's IR-dark ($A_v\sim 100$ mag) and HNC bright ($A_V$ $\gtrsim$ a few mag) radii are sensible.  Using HNC emission directly to calculate mass is dangerous, because, as mentioned above \citep{Jackson2010}, the fractional abundance of HNC is uncertain by as much as three orders of magnitude.   Calculating masses using a filament diameter of $0.01$ pc and a characteristic column density $A_V=80$ mag (corresponding a a volume density of  $10^5$ cm$^{-3}$),  Nessie Classic is $1 \times 10^5$ M$_\odot$, Nessie Extended is $2 \times 10^5$ M$_\odot$ and Nessie Optimistic is $5 \times 10^5$ M$_\odot$.  If one assumes that the envelope traced by the HNC observations of Jackson et al. (2010) for Nessie Classic continues along Nessie's length, then the mass of an $n\sim 500$ cm$^{-3}$ cylinder with diameter 0.1 pc (see Table 1) associated with Nessie would be $5 \times 10^4$ M$_\odot$ for Classic and $3 \times 10^5$ M$_\odot$ for Optimistic.  For the Optimistic case, this mass amounts to 2 millionths of the total gas mass (assuming $\sim 10^{11}$ M$_\odot$ total) of the Milky Way.  

Most of the mass in molecular clouds,  which typically follows a log-normal distribution \citep[cf.][and references therein]{Goodman2009}, is at densities substantially lower than the 500 cm$^{-3}$  we have used for Nessie's outer HNC-emitting regions.   For example, \citet{Battisti2014} find that the Dense Gas Mass Fraction (DGMF) in molecular clouds is of order 10 percent, using the mass in sub-mm-emitting dense dusty cores \citep[from the BGPS Survey,][]{Rosolowsky2010} as the numerator and the full mass of $^{13}$CO-emitting clouds \citep[from the GRS Survey,][]{Jackson2006} as the denominator.  Emission from $^{13}$CO typically traces of order 10 percent of the total molecular mass \citep[cf.][and references therein]{Pineda2008}.  So, correcting for a a 10\% DGMF and $^{13}$CO  tracing at most 10\% of the total H$_2$, in the limiting case that {\it all} dense gas were in Nessie-like structures, the total number of ``Nessie Optimistics" that one could find in the Milky Way is at least 100 times less than the 500,000  listed in Table 1 under ``number to equal mass of Milky Way."  By-eye inspection of Figures \ref{fig:simulationtop} and \ref{fig:simulationedge} gives the impression that roughly 20\% of the very high column density ($A_V\sim 100$ mag) gas will be in easy-to-identify long, straight structures.  So,  a coarse estimate of ``how many Nessies" are detectable in the Milky Way can be based on the ratio ($\sim 500,000$) of the Milky Way's gas mass divided by Nessie's mass traced by HNC, reduced by $\times \sim  0.1$ to account for the  dense gas mass fraction, another $\times \sim 0.1$ to account for the ratio of total molecular to $^{13}$CO-emitting gas, and another at best  $\times \sim 0.2$ to account for the topology in evidence in the Smith et al. simulations.   That leaves us with an estimate that of order many hundreds to thousands of additional Nessie-like features that should be discoverable in the Milky Way.  And, of course, features nearer to us, and oriented more perpendicular to our line of sight, as Nessie is, will be the easiest to find.

To make more robust estimates of how discoverable our Galaxy's bones will be, future work should use simulated observations of galaxy simulations like those presented in \citet{Smith2014}, and/or high-resolution extragalactic observations of spiral galaxies, to statistically model what structures could be observable. 
For reference, the 40 pc resolution of the interferometric state-of-the-art PAWS Survey of CO in the nearby spiral M51 \citep{Schinnerer2013} is not high enough to find Nessie-like structures. Future ALMA observations with resolution of several pc may be able to find evidence for bones by finding their ``envelopes," as traced, for example, by the HNC observations in \citet{Jackson2010}.

\subsection{Can We Map the Full Skeleton of the Milky Way?}
\label{future}

We have long thought that generating an outside-in view of the Milky Way is impossible, almost as inhabitants of Edwin A. Abbott's {\it Flatland} \citep{Abbott1884} residents cannot envision a 3D view of their 2D world.  But, we can escape flatland by realizing that the {\it tiny offset of the Sun above the Milky Way's midplane can give us a tiny, but useful, bit of perspective on the 3D structure of the Milky Way}.  This view is only useful for mapping structure using very high-contrast, very narrow, features like Nessie, because puffier features will overlap too much to be identifiable in such a foreshortened view.  The identification of Nessie as a bone-like feature implies that modern observations are good enough, and features are sharp enough, to identify at least some of our Galaxy's skeleton.

It turns out that Nessie is located in a place where seeing a very long IRDC projected parallel to the Galactic Plane should be  easiest.  Look again at Figure \ref{fig:topview}, and consider Nessie's placement there.  According to the current (data-based cartoon) view of the Milky Way shown in Figure \ref{fig:topview}, Nessie is in the closest major spiral arm (Scutum-Centaurus) to us, along a direction toward, but not exactly toward, the (confusing) Galactic Center.  Nessie's placement there means that it will have a bright background illumination as seen from further out in the Galaxy (e.g. from the Sun), and that it will have a long extent on the Sky as compared with more distant or less perpendicular-to-our-line-of-sight objects.  It is always good when one finds what should be the easiest-to-see example of a new phenomenon first, so we are reassured that Nessie was the first ``Bone" of the Milky Way found.   Now, though, we need to think about how to find Bones not as prominent as Nessie.  There are essentially two approaches.

In one approach, one can conduct a strategic search for long thin clouds, or broken pieces of such, in places on the Sky where they ``should" occur according to  our current understanding of the Milky Way's spatial and velocity structure.  Specifically, using current estimates for the Milky Way's arm's 3D locations, one can draw more velocity-encoded traces like the ones shown in Figure \ref{fig:coloredlines} on the 2D Sky, showing where arms should appear.  With such predictions in hand, one can design algorithms to look for dust clouds elongated (roughly) along those lines, and then examine the velocity structure of the elongated features, as we do in \S \ref{3D}, above.   Of course, one need be flexible about which features one accepts as possible  ``bones," remembering that the model used to draw the expected features on the Sky is the same one being refined!  It is likely that an iterative approach, using the extant Milky Way model as a prior, will succeed in this way.   

Taking a less targeted survey approach, one can search the Sky for long, thin, dense clouds, and assess their relationship to Galactic structure after the fact.  This blind search approach has recently been taken by \cite{Ragan2014} in the first quadrant of the Milky Way, using near and mid-infrared images.  Of the seven new clouds (dubbed ``Giant Molecular Filaments") identified by \citet{Ragan2014} none appears to be tracing spiral arm structures like Nessie.  One  cloud, GMF 20.0-17.9, appears to be a spur off of the Scutum-Centaurus Arm.  \citet{Ragan2014} confirm our claim that Nessie lies in the middle of the Scutum-Centaurus Arm, and they speculate that perhaps identifying bone-like features in the first quadrant is more difficult than in the fourth, where Nessie lies.   Examining Figure \ref{fig:topview}, it is clear that Nessie and other clouds in the Scutum-Centaurus arm would be much more perpendicular to our line of sight, and thus easier-to-detect, in the fourth quadrant than in the first, so we agree with Ragan et al.'s speculation.

In a related approach, a ``connect the dots," strategy for finding long features is also possible. Using the output of exhaustive (automated) searches for high-density peaks, once can look for non-random, long, straightish, patterns in the distribution of those peaks.  For example, in plotting out the \citet{Peretto2009a}  catalog of the positions of 11,000 roundish high-density peaks on the Sky,  long thin IRDCs appear to the eye as strings of easy-to-connect sources.  Nessie itself is comprised of $\sim 100$ Peretto \& Fuller sources superimposed on a slightly-lower-density connecting structure.   It may be possible to use automated structure-finding algorithms, such as Minimum Spanning Trees,  to connect-the-peaks in efforts to identify additional bones of the Milky Way.

Prior to our work on assessing the role of (a lengthened) Nessie in Galactic structure, and Ragan et al.'s 2014 survey, at least two other exceptionally long molecular clouds near the Galactic plane were presented in the literature.   The so-called ``Massive Molecular Filament" G32.02+0.06, studied by \citet{Battersby}, does not appear to be tracing an arm structure. And the 500-pc long molecular ``wisp" discussed by \citet {2013A&A...559A..34L} also does not presently appear directly related to Galactic structure, although it is interesting to think about its past.    As we discuss in \S \ref{spine}, it will likely prove important to model the effects of feedback on the evolution of bone-like star-forming molecular clouds. None of the other long, thin, dense clouds discovered to date are as  elongated {\it and} straight as Nessie. Intuitively, one can guess that over very long time scales, the star formation going on with clouds like Nessie will re-shape them and break them into pieces very hard to recognize as ever having been bone-like.

As extinction, dust emission, and molecular spectral-line maps cover more and more of the sky at ever-improving resolution and sensitivity, it should be possible to map more and more of the Milky Way's skeleton.  New wide-field extinction-based efforts based at first on Pan-Starrs \citep [e.g.][]{Green2014}, and ultimately on Gaia, will be tremendously helpful in these efforts in the coming decade.  Once enough features are identified on the Sky and in 3D space using dust maps, and in position-position-velocity space (using information from spectral-line observations), it should be possible to fit features together into a skeletal pattern of arms, spurs, and more,  in much the same way one solves a jigsaw puzzle by finding the most obvious connections first.  Ultimately, we may have enough examples of ``bones" to apply a machine-learning algorithm that would search for additional pieces of the Milky Way's skeleton based on known examples, in much the same way that \citet{Beaumont2011} used the output of the Milky Way Project's citizen science  effort as input to a Support Vector Machine algorithm that quantifies estimates of the density and characteristics of bubble-like features in the interstellar medium.

\begin{figure}[!htb]
\begin{center}
\includegraphics[width=1\columnwidth]{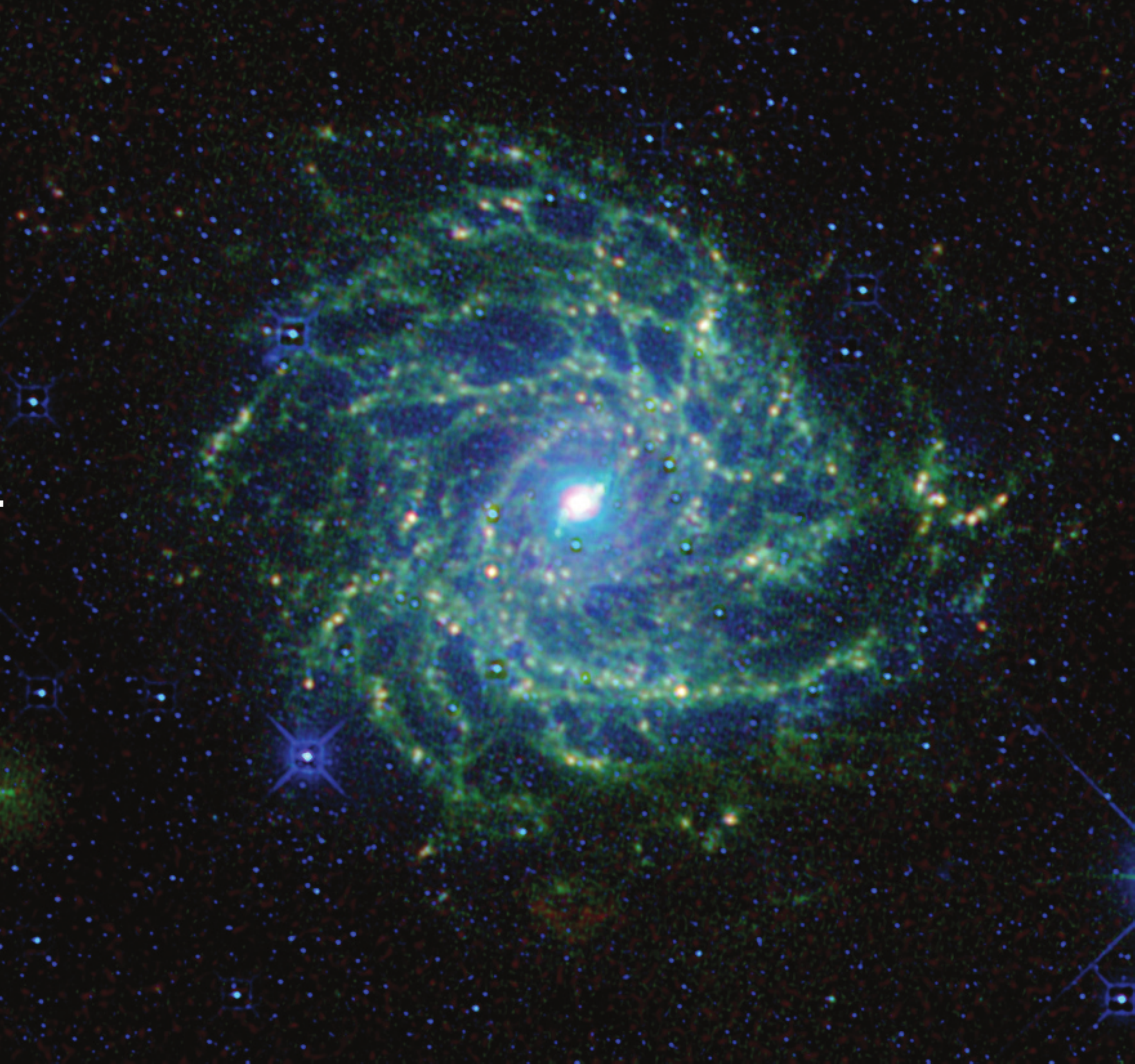}
\caption{\textbf{\label{fig:IC342}}. IC342 as seen by WISE, reproduced from Jarrett et al. 2013. The colors correspond to WISE bands: 3.4 $\mu$m (blue), 4.6 $\mu$m (cyan/green), 12.0 $\mu$m (orange), 22 $\mu$m (red).}
\end{center}
\end{figure}

While many long, thin, IRDCs will likely be identified in the coming years as ``bones" of the Milky Way, it is important to point out that not every long-thin IRDC should be expected to trace a spiral arm.  The image of IC342 in Figure \ref{fig:IC342} and the simulation in Figure \ref{fig:simulationtop}  illustrate that spiral galaxies' structure can include much more than arm-like high-contrast structures, and we can predict that shear and feedback disrupt long  features over time.  This complexity will make piecing together the Milky Way's skeleton from its bones very challenging, but it should be an achievable goal. 

\section{Contributions and Facilities}

\subsection{Contributions}

This paper was a truly a group effort, and the author list includes only
some of the many people who have contributed to it. The entire project
was inspired by a question: ``Is Nessie parallel to the Galactic Plane?,"
asked by Andi Burkert at the 2012 Early Phases of Star Formation
(EPoS) meeting at the Max Planck Society's Ringberg Castle in Bavaria. Three EPoS attendees beyond the author list contributed significant ideas and data to this
work, most notably Steven Longmore, Eli Bressert, and Henrik Beuther. We are grateful to
Cormac Purcell for giving us advance online access to the HOPS data, and
to Mark Reid for generously sharing his expertise on Galactic structure.
The text here was largely written by Alyssa Goodman; the theoretical
ideas come primarily from Andi Burkert and Rowan Smith; and much of the geometrical
analysis was carried out by Christopher Beaumont, Bob Benjamin, and Tom
Robitaille. Tom Dame and Bob Benjamin provided expertise on Galactic
structure, and they created several of the figures shown here. Jens Kauffmann provided expertise on IRDCs, and also was first to point out the potential relevance of the Sun's non-zero height above the Galactic Plane.  Joao Alves provided expertise on the potential for using extinction maps to find more Nessie-like features, and Michelle Borkin was instrumental in early visualization work that led to our present proposals for using the Sun's ``high" vantage point to map out the Milky Way.  Jim Jackson contributed critical expertise on the Nessie IRDC, based both on the 2010 study he led and on unpublished work since.  During revisions to this manuscript, Rowan Smith provided the new simulations referenced here as \citet{Smith2014}.

The article you are reading now was the first to be prepared using a collaborative authoring system called Authorea.  The early drafts of the paper, as well as the (preprint) submitted version, were, and will remain, open to the public, at \href{https://www.authorea.com/249}{this link}. We thank Authorea's founders Alberto Pepe and Nathan Jenkins, and advisors Eli Bressert and Matteo Cantiello, for assistance as the work proceeded. 

The final version of this paper benefitted significantly from responses to the comments of an usually thorough referee, to whom we are grateful.

A.B. acknowledges support from the Cluster of Excellence ``Origin and Structure of the Universe." A.G. and C.B. thank Microsoft Research, the National Science Foundation (AST-0908159) and NASA (ADAP NNX12AE11G) for their support. M.B. was supported by the Department of Defense through the National Defense Science \& Engineering Graduate Fellowship (NDSEG) Program. R.B. acknowledges NASA grant NNX10AI70G.

\subsection {Facilities}
Data in this paper were taken with the following telescopes.  The CO Survey of the Milky Way data \cite{Dame2011} are from the 1.2-Meter Millimeter-Wave Telescope in Cambridge, Massachusetts, USA. NH$_3$ observations of cores \cite{Purcell2012} in and near Nessie are from the Mopra 22-meter telescope near Coonabarabran, Australia. The mid-infrared images of the Galactic Plane used to define Nessie are from NASA's Spitzer Space Telescope, and they were made as part of the GLIMPSE \citep {2003PASP..115..953B, 2009PASP..121..213C} and MIPSGAL \citep{2009PASP..121...76C} Surveys of the Galactic Plane.  


\end{document}